\documentclass[fleqn,10pt]{wlscirep}
\usepackage[utf8]{inputenc}
\usepackage[T1]{fontenc}
\title{Damage due to Ice Crystallization}

\author[1,2,*]{Menno Demmenie}
\author[1]{Paul Kolpakov}
\author[1]{Boaz van Casteren}
\author[1]{Dirk Bakker}
\author[1]{Daniel Bonn}
\author[1]{Noushine Shahidzadeh}
\affil[1]{Van der Waals–Zeeman institute, Institute of Physics, University of Amsterdam, Science Park 904, 1098 XH Amsterdam, The Netherlands. }
\affil[2]{Van ’t Hoff Institute for Molecular Sciences, University of Amsterdam, Science Park 904, 1098XH Amsterdam, The Netherlands}

\affil[*]{M.Demmenie@uva.nl}

\begin{abstract}
The freezing of water is one of the major causes of mechanical damage in materials during wintertime; surprisingly this happens even in situations where water only partially saturates the material so that the ice has room to grow. Here we perform freezing experiments in cylindrical glass vials of various sizes and wettability properties, using a dye that exclusively colors the liquid phase; this allows to precisely observe the freezing front. The visualization reveals that damage occurs in partially water-saturated media when a closed liquid inclusion forms within the ice due to the freezing of air/water meniscus. When this water inclusion subsequently freezes, the volume expansion leads to very high pressures leading to the fracture of both the surrounding ice and the glass vial. The pressure can be understood quantitatively based on thermodynamics which correctly predicts that the crystallization pressure is independent of the volume of the liquid pocket. Finally, our results also reveal that by changing the wetting properties of the confining walls, the formation of the liquid pockets that cause the mechanical damage can be avoided.
\end{abstract}
\begin{document}

\flushbottom
\maketitle
% * <john.hammersley@gmail.com> 2015-02-09T12:07:31.197Z:
%
%  Click the title above to edit the author information and abstract
%
\thispagestyle{empty}

%\noindent Please note: Abbreviations should be introduced at the first mention in the main text – no abbreviations lists. Suggested structure of main text (not enforced) is provided below.

\section*{Introduction}

Most of us have faced at least once in our life the catastrophic situation of having forgotten a bottle of liquid in the freezer, and found it broken the next day.  The freezing of water in confined spaces is an important issue for various industries, including infrastructure maintenance \cite{yang2020pavement,zhang2020influence}, cryopreservation \cite{elliott2017cryoprotectants,chaurasia2020review,grotter2019recent}, agriculture \cite{dalvi2019review,tan2021formation}, and art conservation \cite{tschegg2016environmental,zhang2021study}; in all these cases, frost is known to cause frequent and costly damage. The most common explanation for that is based on the unusual property of water in liquid and solid states in comparison to other materials: In the liquid state, water molecules are more densely packed compared to the their arrangement in the crystalline lattice of ice. Consequently, water undergoes volumetric expansion when transformed into ice, potentially causing high pressures on confining walls and inducing mechanical damage \cite{castell2023frost,vigoureux2018can,akyurt2002freezing,davidson1985photoelastic,dash2006physics,gerber2022stress}. 
However, this explanation falls short as, for instance, mechanical damage is also reported in situations where water does not completely saturate the confined space. This is especially true in situations where the ice is confined in one direction but can expand in another direction; think for instance of a glass bottle in the freezer being only half-filled with liquid, or cases where water pockets are partially saturating  porous materials. One would then expect the ice to be able to grow where there is room for it to expand without inducing mechanical damage. Nonetheless, mechanical damage is indeed observed in these cases. Therefore, the underlying damage mechanism due to freezing is not yet fully elucidated.
In recent years, several studies have been performed on supercooled and salty water droplets (in small volumes) to better understand the early stages of ice formation. Moreover, the role of nanostructured surfaces in the presence of air or oil-impregnated surfaces for either delaying ice formation or inducing dendritic growth in droplets has brought new insights on the ice attachment mechanism during condensation frosting \cite{gandee2023unique,kalita2023microstructure,chu2024interfacial,ma2020atomic}. However, these studies focus on specific microscale configurations and do not clarify the complete mechanism of freezing damage on the macroscale.
\\
\begin{figure*}[t!]
\centering
    \includegraphics[scale=0.21]{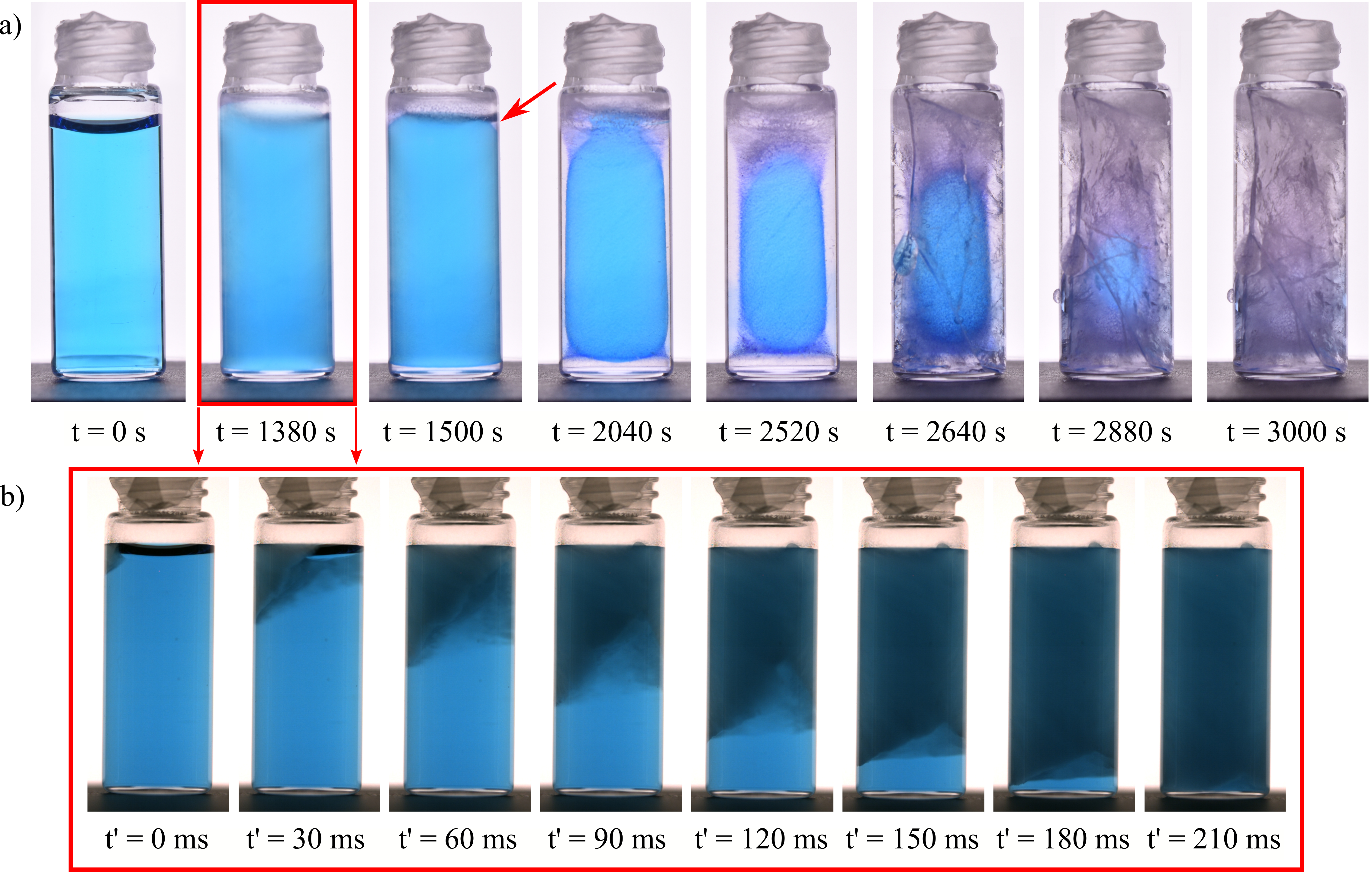}
    \caption{\label{fig:Panel1} Snapshots of key moments in the freezing process of $4$ mL of confined water at an ambient temperature of $-30$ $^\circ$C. Red arrow indicates the location where bulk crystallization initiates, at the contact line of the concave meniscus \textbf{(a)}  Example of two-phase crystallization after sufficient supercooling of the liquid. \textbf{(b)} Detailed chronology of the dendritic crystallization occurring after 1380 seconds in panel a, recorded at a frame rate of 2000 fps. From this sequence of images we obtain a dendritic crystallization velocity of $\sim 15$ cm/s, comparable to literature values \cite{campbell2022dynamics,grivet2022contact}.}
\end{figure*}
\\
In this paper, we investigate ice crystallization in 'open' confinements in cylindrical glass vials of different sizes and wetting properties, partially saturated with water. Our results reveal the major role of the formation of a liquid pocket, or liquid inclusion, within the ice and its subsequent transformation to ice in causing the observed mechanical damage. Our results show that ice nucleation and growth typically initiate at the free surface of a concave air/water meniscus if a hydrophilic glass vial is used. As the ice grows, the meniscus freezes before the bulk which induces the entrapment of a liquid water pocket in the heart of the growing ice. The further transformation of this entrapped liquid volume into ice in the confined space leads to the fracture of both solid materials, i.e., the ice and the glass container. The latter can be understood mechanically and thermodynamically on the basis of an old paper by Gulida and Lifshitz \cite{lifshitz1952theory} who consider the opposite case of a liquid that melts locally within a solid. For a usual solid-liquid transition, the liquid occupies more space than the solid, and so a stress in the solid results from the transformation. 

A hydrophobic treatment of the glass containers suppresses the curved shape of the meniscus to a flat one, shifting the ice nucleation point to the bottom of the glass. As a result, the probability of liquid pocket formation during the total phase transition is drastically reduced. Such treatment could be used as a potential mitigation strategy for freezing-induced damage in confined spaces. Our study also reveals that the degree of supercooling of the liquid before ice nucleation has a significant effect on the kinetics of ice crystallization and the potential for damage. When cooling from the outside, ice crystallization can occur in two stages: first, the formation of fast-growing metastable dendritic ice on the glass wall, followed by the formation of a bulk ice crystal. Interestingly, this two-step process results in a significant number of entrapped air bubbles within the ice, which seem to act as pressure reservoirs and thereby reduce the probability of the container breaking. 

\begin{figure}[h!]
\centering
    \includegraphics[width=0.92\textwidth]{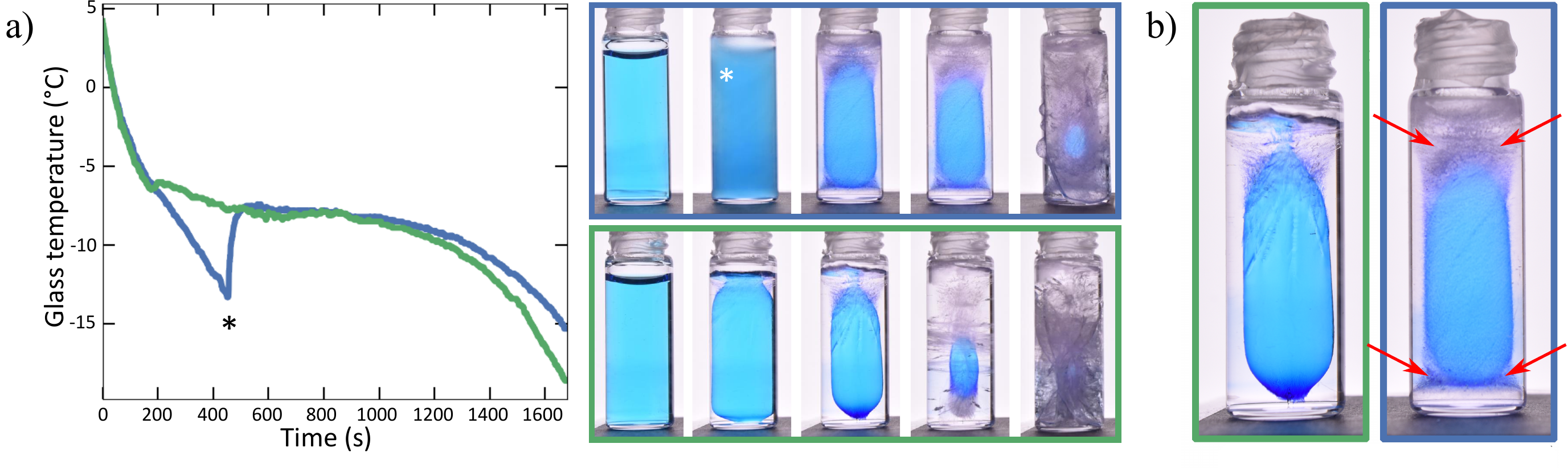}
    \caption{\label{fig:tjump} \textbf{(a)} Measured glass temperature of 2 containers. In blue, a sample that exhibited supercooling ($\Delta T \approx 6^\circ C$) whereas the green sample did not. The occurrence of supercooling goes hand in hand with dendritic crystallization, as indicated by an asterisk. \textbf{(b)} Magnified view of the middle image of the panel in Fig.~(a), at the moment of complete meniscus freezing. The difference in the amount of entrapped air bubbles is clearly visible at the bottom of the liquid inclusions. The sample in the green frame shows transparent bulk ice, whereas the sample in the blue frame, which exhibited dendritic crystallization contains numerous entrapped air bubbles, indicated by the red arrows.}
\end{figure}

\section*{Results}
\begin{figure*}
    \centering
    \includegraphics[width=0.981\textwidth]{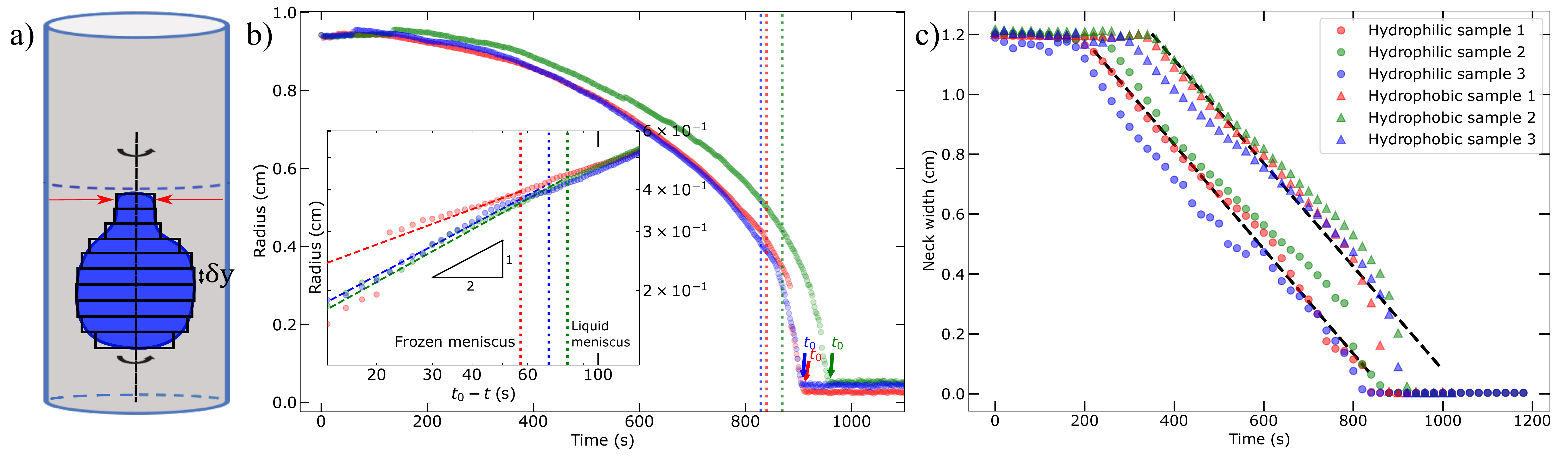}
    \caption{\textbf{(a)} Body of revolution method to determine the volume of the liquid inclusion. $\delta y$ has the thickness of one pixel and therefore, is limited by the camera resolution. Red arrows show the position of the measured neck width, close to the meniscus. \textbf{(b)} Equivalent spherical radius decrease of a liquid inclusion in three experiments in a vial with an inner radius of 7.35 mm. The red symbols are from an experiment in which the glass container fractured at 885 seconds. Dotted vertical lines depict the time at which the meniscus freezes, closing off a route for water to escape. The discontinuity in the red data at $\sim$ t = $885$ sec, is the moment of glass fracture. The unbroken vials show a continuous radius decrease in time (blue and green curves). The inset zooms in on this moment on double logarithmic scale. Herein, the x-axis is inverted (translated to $t_0 - t$, with $t_0$ the moment no more liquid is observed) to investigate the propagation rate of the ice front. Vials that did not fracture follow the expected relation ($r \sim \sqrt{t_0-t}$) for diffusion limited freezing of a liquid pocket. The red data, representing a vial that fractured, deviates from this relation and shows a clear discontinuity at the moment of fracture. \textbf{(c)} Decrease in neck width of three hydrophilically (dots) and hydrophobically (triangles) treated recipients. In hydrophobic bottles, the meniscus freezes at a later stage, although the velocity ($\sim 0.017$ mm/s) remains similar, as indicated by the fit of the slopes with the black dashed lines.} 
        \label{fig:Ice-front-propagation}
\end{figure*}

Fig.~\ref{fig:Panel1}a shows a series snapshots of important moments in the freezing process at $-30$ $^\circ$C. The water was initially at room temperature. After 1380 seconds of cooling at  $-30$ $^\circ$C, the sample first turns turbid due to a rapid dendritic growth of ice. Fig.~\ref{fig:Panel1}b presents a fast sequence of snapshots taken by the rapid camera of the initial stages of the dendritic ice crystallization. In all experiments that exhibit dendritic crystallization, ice dendrites initiate at liquid meniscus. In recent decades, the influence of triple line (water/air/glass contact line) nucleation has been subject of controversy. Droplets freezing on extremely smooth, clean and homogeneously cooled silicon wafer substrates do not favor nucleation at the triple line \cite{gurganus2011fast}. Conversely, in environments with higher surface roughness, nucleation does generally start at the triple line \cite{gurganus2014nucleation,suzuki2007freezing,kar2021faster}. Within the context of this debate, our experiments indicate that the roughness of borosilicate glass vials (measured RMS value of $\sim 1$ $\mu$m) induces triple line nucleation, leading to dendrite growth at the meniscus-glass contact line. After nucleation, dendrites grow with a velocity of $\sim$$15$ cm/s, matching literature values for dendritic ice crystallization on substrates or in confined spaces \cite{campbell2022dynamics,grivet2022contact}. Dendritic crystallization in bulk water has been reported to be significantly slower because thermal diffusion of the latent heat is more efficient in glass than in water \cite{shibkov2003morphology}, indicating that in our experiments, dendrites grow along the glass interface.

When the freezer is opened at the stage where the dendrites have formed, one can still pour out the liquid in the center of the vial and observe that the ice dendrites indeed form a thin layer near the walls. This means that sample is opaque yet remains blue in transmission because there is only a thin layer of dendritic ice formed on the glass interface. If the opaque sample is left at $-30$ $^\circ$C, nucleation and growth of transparent ice follows in a second step and the sample becomes colorless. The latter initiates again at the contact line of the concave meniscus (shown with a red arrow at 1500 s in Fig.~\ref{fig:Panel1}). The water/ice phase transition progresses at the air/liquid meniscus and from the glass wall of the container  towards the central part of the vial; the freezing front of ice can be visualized as the boundary between colorless ice and the blue liquid phase. 

As long as the meniscus is not completely frozen, due to the volumetric expansion of ice, the remaining liquid water can escape towards the air from the opening in the ice/air interface. When the meniscus is completely frozen, the remaining liquid water gets entrapped in the heart of the growing ice. When the trapped liquid pocket freezes, this leads (in this experiment after 2640 s) firstly to the fracture of the ice (having weaker mechanical properties than the glass) followed by the fracture of the glass container.

In total, we observe such a two-step nucleation process of ice in 65 out of 90 untreated glass samples. For the rest of the samples, the growth of transparent ice occurs directly during the phase transition without passing through the first step of metastable dendritic growth. High-speed-camera observations combined with measurements of the glass temperature (Fig.~\ref{fig:tjump}a) confirm the correlation between dendritic crystallization and the supercooling. Dendritic ice forms when freezing occurs at sufficient supercooling, accompanied by a temperature jump. The blue curve in Fig.~\ref{fig:tjump}a indicates a $\Delta T \approx$ $6$ $^\circ$C at the moment dendrites are observed, as depicted by the asterisk. Conversely, the slope of the green curve decreases without a distinct $\Delta T$ for a glass temperature of about $-6$ $^\circ$C. Since no dendrites were observed at that moment we infer that the water is near $T_m$, resulting in bulk crystallization.
%Furthermore, we have examined the dendritic growth speed using the high-speed-camera and found that it matches previous literature with a speed of $\sim$$15$ cm/s.\cite{campbell2022dynamics}. 
%in good agreement with studies showing that the nucleation kinetics is orders of magnitude higher in a wedge (the meniscus' edge) than on flat surfaces  \cite{diao2011role,page2009crystallization}. 

Surprisingly, the emergence of dendritic crystallization preceding bulk crystallization was found to affect the probability of fracture. Statistical analysis depicted in Fig.~\ref{fig:bardiagram}c indicates that out of the 54 glass containers showing dendritic crystallization, $53.7 \%$ fractured. Conversely, among the 36 containers that crystallized directly, $83.3 \%$ fractured. Inspection of the samples reveals a significant increase in the number of entrapped air bubbles for the samples that underwent dendritic crystallization followed by the bulk ice formation (Fig.~\ref{fig:tjump}b). A likely explanation seems to be that the entrapped air bubbles increase the effective compressibility of the ice/air system and in this way attenuate the stress resulting from the volume expansion when the liquid inclusion turns to ice \cite{potter2020study}. This in turn suggests that the presence of air bubbles can play a crucial role in reducing the fracture of confined growing ice.

\subsection*{Entrapment of the liquid inclusion}

To investigate the dynamics of liquid inclusion formation, we tracked the boundary of the blue liquid phase in time. By assuming the inclusion is radially symmetric, we can approximate its volume $V$ by a body of revolution rotating around the vertical axis. The body is divided into multiple slabs one pixel high before each slab is rotated individually around the central axis (Fig.~\ref{fig:Ice-front-propagation}a). Fig.~\ref{fig:Ice-front-propagation}b shows the temporal evolution (extracted every $2$ seconds) of the equivalent radius $R$ of the approximated liquid inclusion volume given by $V =4/3 \pi R^3$. The vertical dashed lines in (b) show the moment when the meniscus was fully frozen ($t_n$). Among the 90 vials subjected to examination, 59 were broken, while the remaining 31 remained intact. 

To examine the propagation rate of the ice front when the liquid pocket has formed, we plot the radius of the inclusion against the time distance ($t_0 - t$) in the inset of Fig~\ref{fig:Ice-front-propagation}b, with $t_0$ the time at which the inclusion is completely crystallized. This allows us to determine the freezing rate from a different experiments giving a different $t_0$ in situations where no liquid can escape, and consequently, only crystallization contributes to the reduction in liquid volume. In the cases where no fracture occurred, we find that the radius of the inclusion scales approximately with the square root of time ($r\sim \sqrt{t_0-t}$), as expected for a process in which the diffusion of latent heat of crystallization is the limiting factor for the freezing speed. To be more precise, a power law fit $r=a(t_0-t)^b$ applied on the blue and green data yields a power of $b = 0.50 \pm 0.013$ and $b = 0.52 \pm 0.015$, respectively. Here, $t_0$ is not a free fitting parameter but the observed time at which all the liquid has crystallized. Remarkably, when fracture does occur, the radius does not obey the predicted shrinking rate in time. We find an exponent of $b = 0.35 \pm 0.023$ for the red data, indicating that ice crystallization before fracture is not only limited only by diffusion, but that the stress buildup also affects the dynamics.
Another notable observation in the red data is the discontinuous decrease in liquid volume at the moment of fracture. Upon fracture, the pressure drops from several hundred MPa (calculation provided below) to ambient pressure. This pressure drop leads to an increase in melting temperature: $\Delta T_{\textrm{m}} \sim 1^\circ$C each 12 MPa \cite{henderson1987melting}. Consequently, crystallization is significantly enhanced at the moment of fracture.

To see what the pressure build-up due to the inclusion is, we need to do the thermodynamics and mechanics of the problem. In 1952, Gulida and Lifshitz \cite{lifshitz1952theory} examined the thermodynamics of local melting. This is mentioned in a footnote in some editions of the Theoretical Physics series of Landau \& Lifshitz, but the original article is in Russian. We add our translation of the paper as an Appendix. The problem posed is that of the local melting of a solid, e.g. by local heating. Since for most substances, the liquid is less dense than the solid phase, local melting (e.g. by a localized heating source within the solid) will lead to a buildup of pressure due to volume expansion, which shifts the melting temperature since the pressure changes the chemical potential of liquid and solid differently. We are here in exactly the same situation since for water the ice is less dense than the liquid, and one can thus use the same calculation to calculate the pressure that the liquid pocket exerts upon the ice when it crystallizes. 
Transposing Gulida and Lifshitz' calculation to our case, the pressure exerted at the liquid-solid interface upon freezing of the trapped water, assuming that the liquid inclusion is homogeneous, is:
\begin{equation} \label{eq:1}
    P=\frac{(4\mu+3k_1)k_2 P_0-4\frac{\delta\rho}{\rho_w}\mu k_1 k_2}{k_1(4\mu+3k_2)},
\end{equation}
where $\mu \sim 2.5-3.5$ GPa is the shear modulus of ice \cite{neumeier2018elastic,chang2021research}, $k_1 \sim 2.1$ GPa and $k_2 \sim 10$ GPa represent the bulk moduli of water and ice, respectively, $P_0 \sim 0.101$ MPa is the ambient pressure, $\delta\rho \sim 90$ kg/m$^\textrm{3}$ the difference in density between ice and liquid water, and $\rho_w \sim 997$ kg/m$^\textrm{3}$ the water density. 
By assuming the water inclusion is an isotropic sphere,  Eq.\,\ref{eq:1} can be applied to calculate the exerted pressure; we find $P \sim 260 \pm 30$ MPa for our experimental parameters. Note the independence of pressure from the enclosed liquid volume. A recent study have found similar results for supercooled droplets frozen from the outside. In cases where the formed crystalline shells are sufficiently thick, the pressure buildup is also invariant with respect to the liquid radius \cite{wildeman2017fast}. 

The next step is to estimate the pressure that our borosilicate vials can support; this can be computed using Griffith's criterion for linear elastic fracture mechanics \cite{griffith1921vi}: 
\begin{equation}\label{eq:2}
    \sigma = \sqrt{\frac{2 E \gamma}{\pi a}}.
\end{equation}
With Young's modulus $E = 62 \cdot 10^9$ Pa, surface energy $\gamma = 55 \cdot 10^{-3}$ J/m$^2$ and an internal flaw diameter $a = 0.3 - 1$ $\mu$m \cite{holmquist2014internal}, we find a  maximum pressure of $\sim 65 \pm 20$ MPa, which is in good agreement with experimental literature \cite{alarcon1994fracture}. Hence, the built up pressure generated by a volumetric increase during freezing exceeds the required pressure to break a glass confinement. The tensile stress of ice, influenced by factors such as temperature and grain size distribution, is in the order of a few MPa \cite{petrovic2003review, currier1982tensile}, approximately one order of magnitude lower than that of borosilicate glass. Therefore, we observe generally in our experiments the first cracks within the ice in the tangential direction (as shown in Fig.~\ref{fig:Panel1}a) before the fracture of the glass vial.

\begin{figure*}[t!]
    \centering
    \includegraphics[clip, trim={0cm 0cm 0cm 0cm },clip, width=1.1\linewidth]{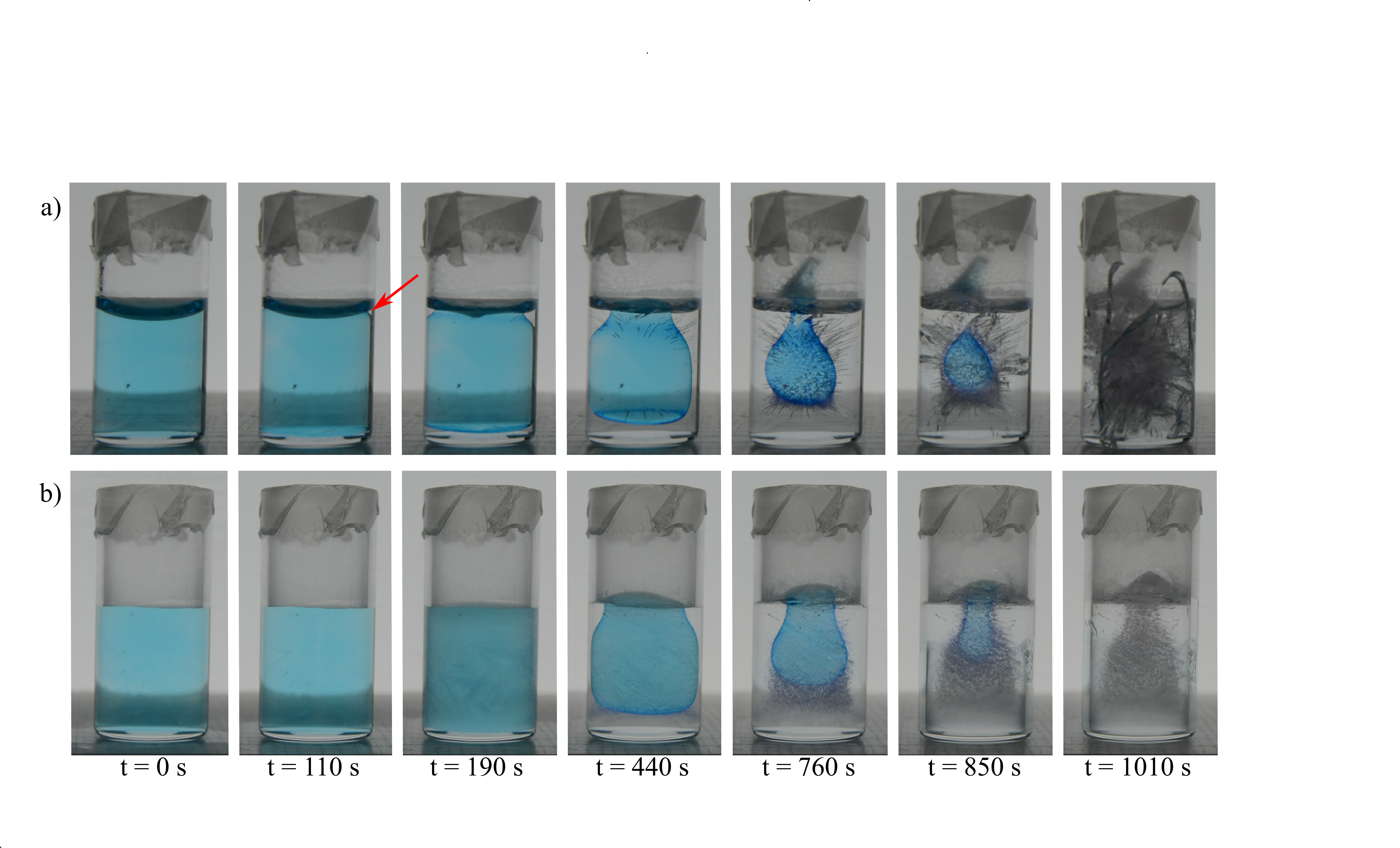}
    \caption{Snapshots of key moments in the freezing process of 2 mL confined water at an ambient temperature of -30 $^\circ$C in two different recipients: \textbf{(a)} hydrophilic, where the liquid gets enclosed and eventually causes fracture in both ice and glass. This sample corresponds to red dots in Fig.~\ref{fig:Ice-front-propagation}b and Fig.~\ref{fig:Ice-front-propagation}c. Red arrow indicates the location where bulk crystallization initiates, at the contact line of the concave meniscus. \textbf{(b)} hydrophobic, demonstrating the absence of a liquid inclusion, therefore no pressure is generated and leaving both the ice and glass intact. This sample corresponds to the red triangles in Fig.~\ref{fig:Ice-front-propagation}c.} 
    \label{fig:hydrophilic-hydrophobic}
\end{figure*}

\subsection*{Wettability of the container and neck closure}

To examine the effect of wettability on the freezing process, we studied the freezing dynamics in three types of glass vials with varying degrees of wettability (supplementary Fig.~\ref{fig:supplementary}) as described in the Methods section. We refer to these as hydrophilic, untreated vials in which water meniscus is concave and hydrophobic confinements with a flat meniscus.
The freezing experiments in these vials reveal the significant role of the location of ice nucleation in the final mechanical damage observed; In both hydrophilic and untreated containers, ice nucleation occurs at the contact line of the concave meniscus. On the other hand, in the hydrophobic recipients, as the 'wedge effect' is almost inexistent for a flat meniscus, nucleation is observed to occur rather at the bottom wall of the glass container. Consequently, while ice growth progresses from the glass wall toward the central part of the container, water can escape for a longer time in the unconfined direction from the unfrozen meniscus (Fig.~\ref{fig:hydrophilic-hydrophobic}). It is interesting to note that the speed of the freezing of the meniscus is found to be constant at approximately 0.017 mm/s, regardless of the wettability, i.e. the shape of the meniscus (fig.~\ref{fig:Ice-front-propagation}c). As a result, there is no entrapped liquid inclusion in hydrophobic vials, which explains the absence of fracturing in the hydrophobic recipients; a total of 22 hydrophobic samples were examined, all of which initiated ice nucleation at the bottom of the glass. Consequently, none of the hydrophobic samples exhibited entrapped liquid inclusions, and no fractures were observed in either the ice or the glass.

\begin{figure}[h!]
\centering
    \includegraphics[width=0.45\textwidth]{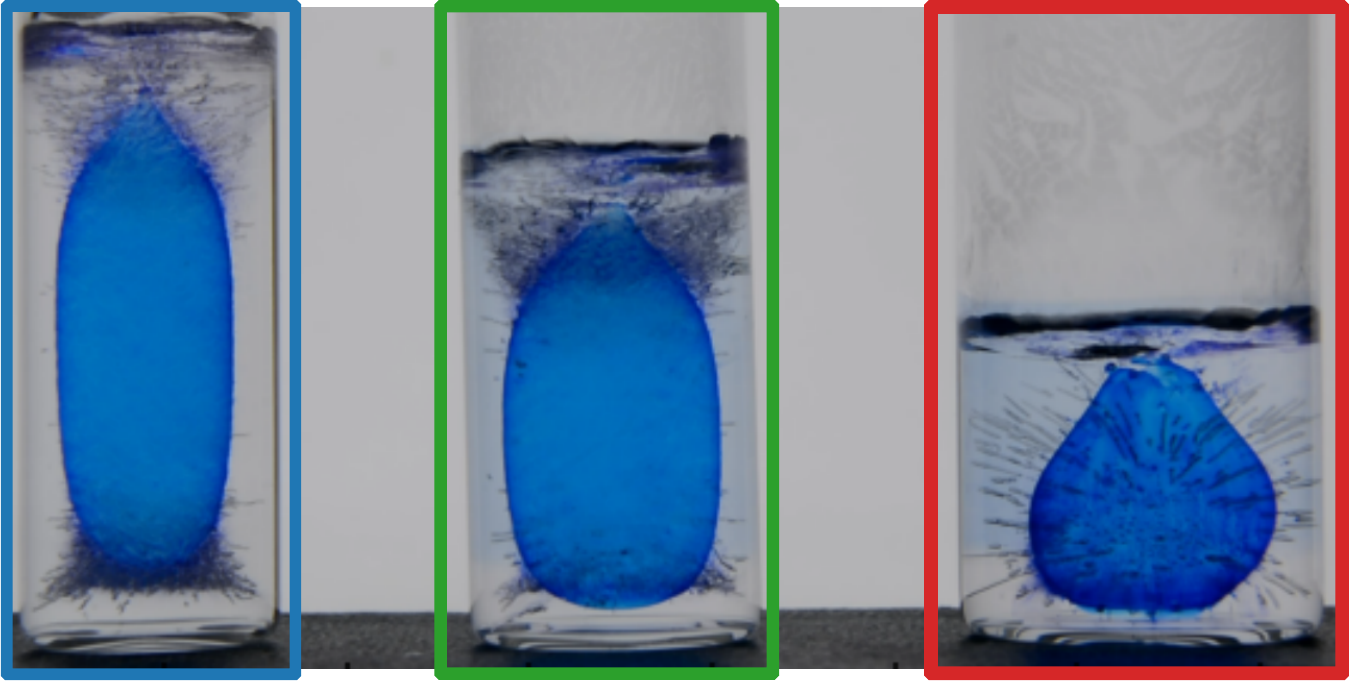}
    \caption{\label{fig:comparison}  Liquid inclusion formation in ice in three glass container used to investigate the effect of the radius-to-volume ratio on the fracture probability. In blue $6.35$ mm, in green $7.35$ mm, in red $9.35$ mm. Images show the moment at which the meniscus freezes completely and the remaining liquid becomes fully enclosed. Colors correspond to the colors in Fig.~\ref{fig:bardiagram}.}
\end{figure}

\subsection*{Radius of the vials}

\begin{figure*}[t!]
\centering
    \includegraphics[width=1\textwidth]{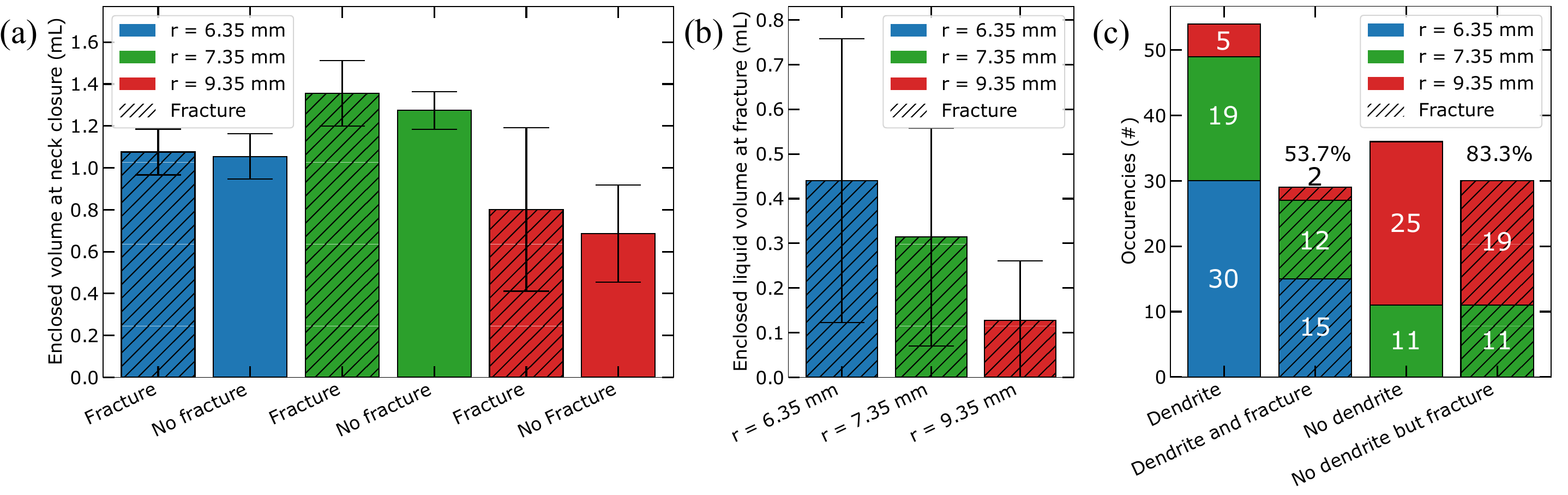}
    \caption{\label{fig:bardiagram} \textbf{(a)} Average entrapped liquid volume at the moment of meniscus closure (cf. Fig.~\ref{fig:comparison}) for glass containers of three different radii. Hatched bars indicate containers that fractured during the experiment whereas non-hatched bars indicate containers that did not. Error bars represent the standard deviation. \textbf{(b)} Average enclosed liquid volume at the moment of fracture. The large error bars indicate that there is no clear threshold in the amount of expansion for fracture to occur. \textbf{(c)} Statistics on the observation of dendritic crystallization. $54$ samples exhibited dendrites of which $53.7 \%$ fractured. $36$ samples did not exhibit dendritic crystallization, leading to a higher fracture probability of $83.3 \%$. Remarkably, all small containers (in blue) underwent dendritic crystallization before bulk crystallization.}
\end{figure*}

The effect of the surface-to-volume ratio of the confinement was also examined by performing experiments on three cylindrical recipients with identical glass thickness ($1$ mm) but varying the inner radius ($6.35$, $7.35$, and $9.35$ mm). The volume of the water was kept constant at $4$ mL. %Measurements were conducted at a temperature of $-30$ $^\circ$C, and repeated 30 times per recipient to generate a statistical basis. 
The smallest glass recipients exhibit a more frequent occurrence of dendritic crystallization prior to bulk ice crystallization (Fig.~\ref{fig:bardiagram}c). This is likely due to the higher surface-to-volume ratio of the smaller containers, leading to a larger degree of supercooling and thus a higher probability of dendritic growth.    
In addition, the shape of the entrapped liquid inclusion beneath the complete frozen meniscus varies in different vials  (Fig.~\ref{fig:comparison}). The wider the cylinder, the more spherical the shape of the liquid inclusion is. The variation in shape results in an average encapsulated liquid volume at meniscus closure of $1.06 \pm 0.11$, $1.34 \pm 0.15$, and $0.77 \pm 0.35$ mL for the containers with $6.35$, $7.35$, and $9.35$ mm radius, respectively. 

Fig.~\ref{fig:bardiagram}a shows the enclosed liquid volume at the moment of meniscus closure for the different size recipients, categorized into two groups: vials that fractured and those that did not. We do not find a statistically significant correlation between the enclosed liquid volume at meniscus closure and the probability of fracture.  In addition, at the moment of fracture, the amount of liquid remaining in the inclusion (Fig.~\ref{fig:bardiagram}b) was quantified. The latter varies widely resulting in large error bars. This again suggests that there is no specific threshold volume expansion value after which fracture occurs, in agreement with the thermodynamics and mechanical arguments presented above. 

\section*{Discussion}

We conducted an experimental investigation to elucidate the freezing kinetics and the mechanism of freezing-induced damage in partially saturated media. We demonstrate that the condition to induce damage is the formation of a liquid inclusion during freezing in bulk ice; when this inclusion freezes,  the $\sim$$9\%$ volume expansion of water during the phase transition to ice causes the damage. By the addition of methylene blue to water, we have been able to visualize ice front propagation and the kinetics of liquid inclusion formation. The subsequent pressure buildup upon its conversion to ice was calculated to reach $\sim$260 MPa, which is indeed large enough to break the glass container walls. Consistent with the theoretical prediction that pressure is independent of the volume of the liquid inclusion, we observe no significant difference in inclusion size and fracture probability (Fig.~\ref{fig:bardiagram}a). Furthermore, the wide variation in enclosed liquid volume present at the moment of fracturing (Fig.~\ref{fig:bardiagram}b) highlights the lack of correlation between entrapped liquid volume and fracture risk.

We demonstrate that hydrophobic coatings can prevent the formation of liquid inclusions by increasing the contact angle of the meniscus, which delays crystallization at the water/air interface. This finding aligns with recent work by Bi et al., who showed that geometric constraints such as those imposed by wedges, limit the movement of water molecules and enhance ice crystallization by aligning the molecules with the ice structure \cite{bi2017enhanced}. This effect is similar to lowering the entropy-related component of the free energy barrier. The authors also propose that the traditional concept of lattice matching between a nucleation site and a crystalline lattice should be expanded to include unconventional surface topographies. Our findings support this idea: the water/air meniscus, with its contact angle against the glass wall, acts as a geometric constraint similar to wedges, which  promotes ice nucleation in conventional hydrophilic glass containers.

We also found that the size of the confinement can affect the freezing dynamics and crystallization process. Smaller glass containers induces larger supercooling; these exhibit a more frequent occurrence of metastable dendritic crystallization prior to bulk ice crystallization. The presence of dendritic ice growth at the glass-liquid surface was found to negatively correlate with the fracture probability. This is likely due to of large amount of entrapped air bubbles that appeared in the bulk ice crystal for samples that underwent first the metastable dendritic nucleation. The formation of bubbles during freezing is a complex process involving the interplay between metastable crystal dissolution and the progression of the freezing front. Recent studies have examined how bubbles form and become trapped withing growing ice crystals \cite{dao2022entrapment,li2022effects}: As the temperature drops, dissolved air is expelled from the ice crystal lattice and becomes entrapped either within the structure itself or in the surrounding liquid of the rapidly growing dendritic crystals. Since the dendritic crystal represents a metastable state that persists until conditions change, the nucleation and growth of a stable ice crystal eventually takes over. During this dissolution process, local changes in pressure and surface tension occur, which facilitate cavitation and further bubble formation \cite{nicoud2018estimation, katsuno2023conversion}.  
As a consequence, the compressive strength of ice decreases, leading to the absorption of the stress generated by the expansion of the liquid inclusion while converting to ice. The recipe to prevent catastrophies in your freezer is therefore to have small and hydrophobic bottles.

\section*{Methods}

The experimental setup consisted of a climate chamber (Weiss KWP 240) with a window that allowed for observation. For the freezing experiments the temperature was mostly maintained at a fixed value of $-30^\circ$C. Glass vials of different sizes were subjected to a rigorous cleaning process with ethanol and deionized water (Millipore, $\rho \approx 18.2$ M$\Omega\cdot$cm) to ensure the removal of any contaminants. Following the cleaning process, the vials were filled with 4 mL of deionized water, which had been degassed in a vacuum chamber for 2 hours. To visualize the propagation of the ice front, we add 8 mg/L methylene blue, a blue dye that is very soluble in water. As with any impurity, it gets expelled by the crystal lattice during the liquid/solid  phase transition. And even if it gets entrapped in the ice due to its rapid growth, it forms dimers and tetramers that are not colored; therefore it exclusively probes the liquid phase in blue \cite{heger2005aggregation, fernandez2020visible, vesely2023impact}.
which allows for the unambiguous detection of the liquid /ice interface during freezing experiments. We checked that the addition of methylene blue did not in any way influence the crystallization location or dynamics. Prior to being placed inside the climate chamber, the vials although open were covered with Parafilm to minimize possible impacts of dust and changes in humidity. To investigate the effect of the geometry, we used glass vials with three inner radii of 6.35, 7.35, and 9.35 mm. The glass thickness of all vials was 1 mm as measured with a confocal profilometer (Keyence VK-X1100). 
Glass containers with different wetting properties were used to study the effect of surface hydrophobicity or hydrophilicity on the freezing  process. To have a hydrophobic surface, glass vials were washed with water and isopropanol, dried, cleaned by plasma treatment (30 s), and submerged into a solution of toluene and trichlorooctylsilane (from Tokyo Chemical Industry, 1$\%$ vol) for 15 minutes. They were subsequently generously rinsed with isopropanol and water. For a hydrophilic surface, after the washing step, glass vials  were treated with an air plasma for 1 minute. The effect of the treatments results in different shapes of the liquid meniscus (flat or concave) since it affects the contact angle of the air/water meniscus on the glass wall, as shown in supplementary Fig.~\ref{fig:supplementary}. Contact angles of $90\pm 3.6^{\circ}$, $79\pm 2.1^{\circ}$, and $68\pm 2.6^{\circ}$ were obtained for hydrophobic, untreated, and hydrophilic containers, respectively. A Nikon D5200 camera and a Phantom TMX 7510 high-speed camera were used for taking temporal images of the water/ice phase transition during freezing.
Thirty measurements were performed for each vial size or wettability property to allow for a statistical analysis in terms of the probability of fracturing. In total more than 120 experiments were been performed for this study.

\section*{Data availability}
All data are available in the main text or the supplementary materials, for further needs contact corresponding author.

%\bibliography{sample}
\providecommand{\noopsort}[1]{}\providecommand{\singleletter}[1]{#1}%

\section*{Acknowledgements}

The authors thank Benoit Estienne and Vladimir Gritsev for access to [1] I. Lifshitz and L. Gulida, On the theory of local melting.
In \textit{Dokl. Akad. Nauk SSSR}, Vol. 87, 377–380 (1952).
This work is financially supported by NWO Projectruimte Grant No. 680-91-133.

\section*{Author contributions statement}
N.S. conceived the experiments,
M.D., P.K., B.v.C. and D.Ba. conducted experiments,
M.D., P.K., B.v.C., D.Ba., D.Bo. and N.S. analysed results,
M.D., P.K., D.Bo. and N.S. wrote original draft,
D.Bo. and N.S. supervised the process. All authors reviewed the manuscript. 

\section*{Competing interests}
The authors declare no competing interests.

\section*{Additional information}

\textbf{Correspondence} and requests for materials should be addressed to M.D.

\section*{Supplementary Information}
\begin{figure}[h!]
  \centering
  \includegraphics[scale=0.35]{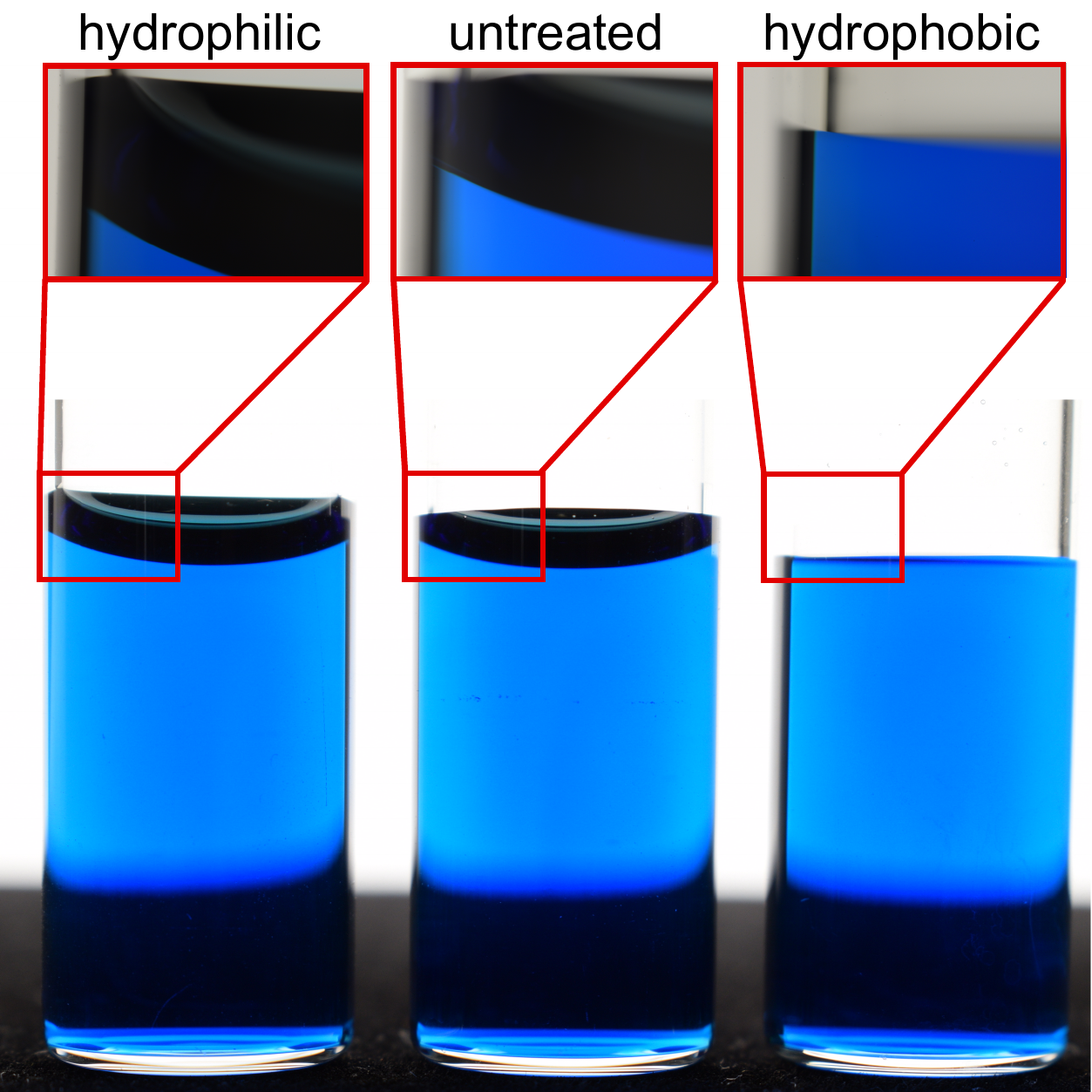}
  \caption{\label{S2} The effect of glass wettability treatments on the meniscus of water inside glass vials. The contact angle of the menisci of hydrophilically treated ($\theta_{hpl}$), untreated ($\theta_{ut}$) and hydrophobically treated ($\theta_{hpb}$) recipients yielded values of $68~\pm~2.6^{\circ}$,  $79~ \pm~2.1^{\circ}$ and $90~\pm~3.6^{\circ}$ at $t=0$, respectively. }
  \label{fig:supplementary}
\end{figure}
\newpage

\newpage

\section{Appendix}

\noindent What follows below is a translation of the iconic paper "On theory of local melting" written by I. Lifshitz and L. Gulida.
\\\\
When considering the process of melting solids, it is usually assumed that pressures and temperatures remain constant along the solid phase. In the case of a temperature gradient or an inhomogeneous stress state in a solid, one can expect the appearance of melting centers inside the solid phase with their subsequent development into local regions of the liquid phase. Let, for example, a heat source be located inside a solid body, and such a temperature is maintained at its boundary that excludes the possibility of melting from the surface. However, in some areas inside the solid phase, the temperature can be so high that it becomes possible for the appearance of local melting centers that arise in a fluctuation way.\\
\\
Assuming the size of the nuclei to be small compared to the distances at which the temperature $T$ or stresses $\sigma_{ik}$ change noticeably, it will suffice to consider the appearance of a single liquid nucleus in an unlimited solid phase homogeneous with respect to $T$ and $\sigma_{ik}$. The choice of the potential $\phi$, which describes the state of the system, is, to a certain extent, arbitrary, since the nature of the conditions specified on the surface, which is infinitely distant relative to the nucleus, is insignificant.\\
\\
One can, for example, set a uniform external pressure $P_0$ on this surface and describe the system by the thermodynamic potential $\Phi$ or fix the volume of the body, characterizing the state of free energy $F$. Then, $\Delta \Phi$ in the first case and $\Delta F$ in the second case will coincide. In what follows, for definiteness, we will assume that the external pressure is fixed.\\
We represent $\Delta \Phi$ as the sum of three terms:\
\begin{align}
    \Delta\Phi = \Delta\Phi_{heat} + \Delta\Phi_{vol} + \alpha\sigma,
\end{align}
$\Delta\Phi_{heat}$ is due to the change in phase stage at $P_0$; $\Delta\phi_{vol}$ is due to the change in strain energy and volume of the system; $\alpha$ is the surface tension coefficient at the interface between the solid and liquid phases; $\sigma$ is the nucleation surface area.\\
Denoting by $T$ the actual temperature of the solid phase, and by $T_0$ the melting temperature at $P_0 = 0$ and expanding $\Delta\Phi_{heat}$ in a series in steps $T-T_0$, we obtain:
\begin{align}
    \Delta\Phi_{heat} = - \frac{q}{T}(T-T_0)v,
\end{align}
Where $q$ is the heat of fusion at $T=T_0$, calculated per unit volume, and $v$ is the volume of the liquid nucleus.\\
We write $\Delta\Phi_{vol}$ as follows:
\begin{align}
    \Delta\Phi_{vol}=\frac{P^2}{2k_2}v+\int_{V-v}wdv+P_0\Delta V-w_0V,
\end{align}
Where $k_1$ and $k_2$ are volume compression moduli of the solid and liquid phases; $P$ the pressure occurring at the surface of the liquid nucleus; $w_0 = P^2_0/2k_1$ and $w$ are energy densities of deformation of solid phase in initial and final states; $V$ is initial volume of system; and $\Delta V$ is the change in volume of the whole system caused by the formation of the nucleus.\\
$w$ and $P$ at given values of $P_0$ and volume $v$ depend on the shape of nucleus surface. This shape must be determined from the condition of a minimum potential $\Phi$. Thus, in order to actually deduce $\Delta \Phi$ with equation 1, a complex variational problem has to be solved.\\
The pressure $P$ on the surface of the nucleus is determined from the equation
\begin{align}
    P=-k_2\left(\frac{v_l(P)-v_l(0)}{v_l(0)}\right) = -k_2\left(\frac{\delta v}{v_{s}(0)}+\frac{\delta \rho}{\rho_s}\right)
\end{align}
Here $v_l$ and $v_s$ are the volumes of particles forming the nucleus in the liquid and solid phase at the corresponding pressures; $\rho_l$ and $\rho_s$ are the densities of the liquid and solid phases; $\delta \rho = \rho_l - \rho_s$. It is assumed that $\delta \rho \ll \rho_s$. $\delta_v = \delta_v(P)$ can be interpreted as a change in the volume of the cavity occupied by the liquid nucleus when a uniform pressure $P$ is applied to its boundary. $\delta_v$ is determined by solving basic equations of elasticity theory under the following boundary conditions:\\
\begin{subequations}
    \begin{align}%{2}
        &\sigma_n = - P \textrm{ at the nucleus surface,}\\
        &\sigma_n = - P_0 \textrm{ at the infinitely distant surface,}
    \end{align}
\end{subequations}
where $\sigma_n$ is the normal stress.\\
In an isotropic solid phase under uniform external pressure, the liquid nucleus will be confined to a sphere. The radial displacement $u_R$ satisfying equation 5 is:
\begin{align}
    u_R = \left(a+\frac{b}{R^3}\right)R, a = \frac{-P_0}{3k_1}, b = \frac{(P-P_0)R_0^3}{4\mu},
\end{align}
with $R_0$ the nucleus radius and $\mu$ the shear modulus.\\
Using equation 3 and 4 and we find that
\begin{align}
    P=-k_2\left(\frac{3u_{R_0}}{R_0} + \frac{\delta \rho}{\rho_s}\right) =\frac{(4\mu+3k_1)k_2 P_0-4\frac{\delta\rho}{\rho_w}\mu k_1 k_2}{k_1(4\mu+3k_2)},\\
    \Delta \Phi = \left(\Psi - \frac{q\Delta T}{T_0}\right)v+4\pi R_0^2\alpha
\end{align}
Where
\begin{align}
    & \Psi = \Psi(P_0) = \frac{(k_2-k_1)(4\mu+3k_1)P_0^2}{2k_1^2(4\mu+3k_2)} - \nonumber\\
    & \frac{k_2(4\mu+3k_1)\delta \rho P_0}{k_1(4\mu+3k_2)\rho_s}+\frac{2\mu k_2}{4\mu + 3k_2}\left(\frac{\delta \rho}{\rho_s}\right)^2.
\end{align}
Using expression 8, it is possible to determine at which external pressure local melting is possible if the temperature is fixed.\\
For a given $\Delta T = (T-T_0)$ the possible values of $P_0$ are determined by the condition:
\begin{align}
    \Psi(P_0)-\frac{q}{T_0}\Delta T \ll 0.
\end{align}
The curves corresponding to equation 10 are parabolas directed upwards for $k_2 < k_1$ and downwards for $k_2 > k_1$.\\
If $k_2 < k_1$, then local melting is possible at pressure $P_0$, or greater than $P_2$, or less than $P_1$, where $P_1$ and $P_2$ are roots of equation 10.\\
if $k_2 > k_1$, then local melting is possible at $P_1 < P_0 < P_2$, when $\delta \rho < 0$ and $\Delta T \rightarrow 0$ roots of $P_{1,2}$ can be negative. In this case, local melting is possible only under conditions of all-round stretching. If $\Delta T$ is negative and large, then the roots of equation 10 will be complex. At such temperatures, local melting is generally impossible. \\
The temperature $T$, corresponding to the beginning of local melting, will also be determined from equation 10, in which $P_0$ should be considered fixed:\\
\begin{align}
    \Delta T_{loc} = \frac{\Psi}{q}T_0 
\end{align}
Expanding in the ratio
\begin{align}
    \Phi_{liq}(P_0,T) - \Phi_{sol}(P_0,T) = 0
\end{align}
in potentials of each of the phases in a series in powers of $T-T_0$ and $P_0$, we obtain the dependence of the usual melting temperature on pressure:
\begin{align}
    \Delta T_{usual} = \frac{T_0}{q}\left[\frac{k_2-k_1}{2k_1k_2}P_0^2 - \left(\frac{\delta \rho }{\rho_s}\right)P_0 \right].
\end{align}
Curves 11 and 13 are parabolas with tangent points at $P_0 = \frac{k_1 k_2 \delta \rho}{\rho(k_2-k_1)}$, and $\Delta T_{loc} - \Delta T_{usual} > 0$.\\
Thus, local melting occurs under the conditions of overheating of the solid phase (at fixed $P_0$ and $T$). The dimensions of the liquid nucleus are determined from the equation $\frac{d \Delta \Phi}{d R_0} = 0$ where,
\begin{align}
    R_0= \frac{-2 \alpha}{\Psi - \frac{\Delta T}{T_0}q}.
\end{align}
Using equation 8 and 14 we find that the probability of the formation of a nucleus:
\begin{align}
    W \sim  e^{\Delta \Phi / kt} = exp\left[\frac{-2 \pi (2 \alpha)^3}{3kT(\Psi-\frac{\Delta T}{T_0}q)^2}\right].
\end{align}
Expressions 8 and 14 for $\Delta \Phi$ were obtained under the assumption that the deformations are purely elastic. If $k_2 - k_1$ or $\delta \rho$ become large enough, then plastic deformations will appear in the solid phase.\\
Using equation 3 and 4 and the well-known solution of the problem of equilibrium of an elastic-plastic spherical shell, it is possible to obtain the equation for determining $P$:
\begin{align}
    \left(\frac{k_2}{k_1} - 1\right)P \mp \frac{k_2}{\alpha} e^{\frac{\sqrt{3}}{2}\frac{P-P_0}{k_2}-1}-k_2\frac{\delta \rho}{\rho_s} = 0,
\end{align}
and $\Delta \Phi_{usual}$ will be equal to:
\begin{align}
    & \Delta \Phi_{usual} = \frac{P^2}{2k_2} - \frac{P_0^2}{2k_1}+\frac{3}{8 \mu}(P_1^2-P_0^2)\nonumber+\\
    & \frac{P_1^2-P_0^2}{2k_1}+\frac{1}{\alpha_1}\left(\pm P - \frac{2k_3}{\sqrt 3}\right) \cdot \nonumber \\
    & exp\left(\pm \frac{\sqrt 3}{2}\frac{P-P_0}{k_3}-1\right)\mp \frac{P_0}{\alpha_1} -\frac{\Delta T}{T_0}q,
\end{align}
where $k_3$ is the yield point; $\alpha_1 = \frac{2\sqrt3 \mu k_1}{k_3(4\mu+3k_1)}$, $P_1 = P_0 \pm \frac{2k_3}{\sqrt(3)}$ is the value of $P$ at which a plastic interlayer appears. The upper sign in 16 and 17 should be taken at $P-P_0>0$; the lower one at $P-P_0<0$. It follows from the consideration of relation 17 that for $k_2 = k_1$ and $\delta \rho \neq 0$ i.e. $\Delta T_{loc}-\Delta T_{usual} > 0$ and taking into account the plastic character of deformation, local melting begins under overheating conditions. For metals, relation 12 gives overestimated temperatures for the onset of local melting. It is possible to indicate other particular cases when $\Delta T_{loc}-\Delta T_{usual} > 0$. Without exact knowledge of the root of equation 16, however, it is impossible to assert that this inequality will be satisfied or any values of $k_2$, $k_1$ and $\delta \rho$.\\
In conclusion, we note that, from the point of view of the hypothesis of local melting, it seems possible to interpret the experiments of S. E. Khaikin and N. P. Binet, who studied the process of pressing tin rods under conditions of an artificially created temperature gradient. In polycrystalline samples, melting was observed, starting inside the solid phase, while in single crystals of tin k, overheating of $1.5 - 2^\circ$C was observed.

\end{document}